\documentclass[reprint,twocolumn,superscriptaddress,showkeys,
nofootinbib,notitlepage,amsmath,amssymb,floatfix]{revtex4-2}

\pdfoutput=1
\usepackage{latexsym,amsmath,amssymb,lmodern,float,url}
\usepackage{natbib}
\usepackage{color}
\usepackage{multirow}
\usepackage{bm}
\usepackage{array}
\usepackage[normalem]{ulem}
\usepackage{cleveref}
\usepackage{xcolor}

\usepackage{tikz}
\usetikzlibrary{datavisualization}
\usetikzlibrary{datavisualization.formats.functions}
\usetikzlibrary{plotmarks}

\usepackage{mathtools}
\def\de{\delta^{\vphantom{1}}}
\def\bde{{\bar\delta}}

\def\ccss{{c\bar{c}s\bar{s}}}

\def\ccqq{{c\bar{c}q\bar{q}^\prime}}

\def\h3{{\displaystyle{\frac 3 2}}}

\newcommand{\bbar}{\overline}

\def\schro{Schr\"odinger~}

\begin{document}
\title{Diabatic Dynamical Diquark Bound States: Mass Corrections and Widths}
\author{Richard F. Lebed}
\email{Richard.Lebed@asu.edu}
\author{Steven R. Martinez}
\email{srmart16@asu.edu}
\affiliation{Department of Physics, Arizona State University, Tempe,
AZ 85287, USA}
\date{April, 2024}

\begin{abstract}
Using the diabatic formalism, which generalizes the adiabatic approximation in the Born-Oppen\-heimer formalism, we apply well-known Hamiltonian methods to calculate the effect of open di-meson thresholds that lie well below the mass of elementary $c\bar c q\bar q^\prime$, $c\bar c s\bar s$, and $c \bar c q \bar s$ tetraquark bound states.  We compute the resulting mass shifts for these states, as well as their decay widths to the corresponding meson pairs.  Each mass eigenstate, originally produced using a bound-state approximation under the diabatic formalism, consists of an admixture of a compact diquark-antidiquark configuration (an eigenstate of the original dynamical diquark model) with an extended di-meson configuration induced by the nearest threshold.  We compare our results with those from our recent work that employs a scattering formalism, and find a great deal of agreement, but also comment upon interesting discrepancies between the two approaches.
\end{abstract}

\maketitle

\section{Introduction}

With over two decades of progress since the discovery of the famous charmoniumlike state $\chi_{c1}(3872)$~\cite{Choi:2003ue}, including the discovery of an impressive 60+ other heavy-quark exotic candidates, it remains conspicuous that no single theoretical framework has emerged to accommodate every state and its properties.
%~\cite{Lebed:2016hpi,Chen:2016qju,Hosaka:2016pey,Esposito:2016noz,Guo:2017jvc,Ali:2017jda,
%Olsen:2017bmm,Karliner:2017qhf,Yuan:2018inv,Liu:2019zoy,Brambilla:2019esw,Chen:2022asf,Lebed:2022vfu}.
Indeed, multiple approaches have been proposed over the years, and it is quite possible that the most effective description of exotics involves the simultaneous use of more than one physical picture.  Taking $\chi_{c1}(3872)$ as an example, with a measured mass from the Particle Data Group (PDG)~\cite{ParticleDataGroup:2022pth} that satisfies
\begin{equation}
m_{\chi_{c1}(3872)} - m_{D^0} - m_{D^{*0}} = -0.04 \pm 0.09 \ {\rm MeV} ,
\label{eq:XBE}
\end{equation}
it is completely natural to describe the state as a loosely bound hadronic molecule of $D^0$ and $\bar D^{*0}$ (plus charge conjugate).  Indeed, this type of molecular picture has a long history, going back nearly 30 years prior to the discovery of $\chi_{c1}(3872)$~\cite{Voloshin:1976ap,DeRujula:1975qlm}.  In addition, a number of other heavy-quark exotic states, such as $Z_c(3900)$ or $Z_b(10610)$, lie close ($< \! 15$~MeV) to a nearby di-hadron threshold.  These $Z$ states however, lie slightly \textit{above\/} the nearest di-meson threshold, discouraging at least a naive interpretation as a hadronic molecule, and requiring a more sophisticated interpretation in terms of scattering theory and the quantum field-theoretical effects induced by the threshold states.  Furthermore, the state $\chi_{c1}(3872)$ has been measured to have a significant ($> \! 10\%$) decay branching fraction to charmonium (including radiative modes), which points towards the state having some appreciable short-range component in its wave function.  The most obvious physical configuration that could provide this component is the currently unobserved conventional charmonium state $\chi_{c1}(2P)$, and in fact including such a component in admixture with a $D^0$-$\bar D^{*0}$ molecule has long provided a leading proposal for the structure of $\chi_{c1}(3872)$~\cite{Suzuki:2005ha}.  However, an additional compact-configuration  candidate is available: a color-antitriplet diquark $\de \! \equiv \! (cq)_{\mathbf{\bar3}}$ plus a color-triplet antidiquark $\bde \! \equiv \! (\bar c \bar q)_{\mathbf{3}}$, with $q=u$ for $\chi_{c1}(3872)$.  One specific framework, the \textit{dynamical diquark model}~\cite{Brodsky:2014xia,Lebed:2017min}, has been thoroughly developed to describe the spectroscopy of heavy-quark exotics in terms of $\de$-$\bde$ states~\cite{Lebed:2017min,Giron:2019bcs,Giron:2019cfc,Giron:2020fvd,Giron:2020qpb,Giron:2020wpx,Giron:2021sla,Giron:2021fnl}. 

In contrast to molecular and threshold-effect descriptions, the dynamical diquark model has the advantage of being able to describe the existence of exotic states such as $Z_c(4430)$ that lie far from any relevant threshold.  Additionally, the model includes effects from spin- and isospin-dependent interactions, thereby generating complete multiplets---in multiple flavor sectors---that include fine structure~\cite{Giron:2019cfc,Giron:2020fvd,Giron:2020qpb,Giron:2020wpx,Giron:2021sla,Giron:2021fnl}.  The most recent developments in this model include the incorporation of mixing between $\de$-$\bde$ and di-meson configurations in order to produce bound~\cite{Lebed:2022vks} and scattering~\cite{Lebed:2023kbm} states.  The extension of the dynamical diquark model to accommodate these threshold states is performed using the \textit{diabatic formalism}, originally developed in atomic physics~\cite{Baer:2006}, which was first applied to heavy-quark exotics in Ref.~\cite{Bruschini:2020voj} in the analogous context of mixing conventional quarkonium with threshold effects.  This extended ``diabatic'' formalism (which we describe in greater detail below) is the rigorous generalization of the Born-Oppenheimer (BO) approximation~\cite{Born:1927boa}, which serves as part of the foundation for the dynamical diquark model~\cite{Lebed:2017min}.

The hidden-charm states in Ref.~\cite{Lebed:2022vks} (subsequently reproduced in the scattering formalism of Ref.~\cite{Lebed:2023kbm}) are obtained using a bound-state approximation, which decouples open thresholds from mixing with the mass eigenstates.  While this approximation is effective for states that lie below or just above a threshold, those whose eigenstates lie far above an open threshold are more appropriately handled as scattering resonances, which in turn is the focus of Ref.~\cite{Lebed:2023kbm}.  Alternately, it is possible to expand upon the bound-state approach in the diabatic formalism by incorporating open thresholds via the methods of Ref.~\cite{Bruschini:2021cty}, which employs coupled-channel techniques previously used in the early days of charmonium studies~\cite{Eichten:1978tg}, and earlier still in atomic physics~\cite{Fano:1961zz}.  Using this approach allows for a comparison with results from the scattering calculations, as discussed below.  To perform further tests, we also compute $1^{--}$ $c\bar c s\bar s$ and $1^-$ $c\bar c q\bar s$ states both in the scattering formalism of Ref.~\cite{Lebed:2023kbm} (where they did not appear) and in the current approach.

This paper is organized as follows.  In Sec.~\ref{sec:DDD Model} we briefly review the original dynamical diquark model, subsequently modified to incorporate the diabatic formalism.  Section~\ref{sec:MS} reviews the explicit methods applied in Ref.~\cite{Bruschini:2021cty} to be used in the diabatic dynamical diquark model ({\it i.e.}, the inclusion of explicit mixing with open di-hadron thresholds), in order to compute mass corrections and decay widths for the eigenstates of Refs.~\cite{Lebed:2022vks,Lebed:2023kbm}.  In Sec.~\ref{sec:Results} we discuss the significance of the resulting mass corrections and decay widths retrieved via those methods, and in Sec.~\ref{sec:Conclusions} we summarize our results, in addition to providing future directions for expanding upon this work.

\section{The (Diabatic) Dynamical Diquark Model}\label{sec:DDD Model}

The dynamical diquark model provides a general mechanism for describing the formation of both tetraquark~\cite{Brodsky:2014xia} and pentaquark~\cite{Lebed:2015tna} states (and can be extended to other multiquark configurations as well~\cite{Brodsky:2015wza}); here, for simplicity we confine our attention to just the tetraquark case.  For any process that generates the valence-quark content $Q  \bar Q  q \bar q^{\, \prime}$, where $Q$ is a heavy quark, the quark pair ($Qq$) and the antiquark pair ($\bar Q \bar q^{\, \prime}$) are produced in an appreciable fraction of the events such that their internal momenta are small relative to that between the pairs, allowing for the respective pairs to nucleate into compact diquarks $\de$ and $\bde$, respectively.  Being bound by confinement, $\de$ and $\bde$ must eventually transfer the bulk of their available kinetic energy to the color flux tube (string) connecting them.  This ``stretched" configuration has difficulty hadronizing into a di-meson pair $(Q\bar q^{\, \prime})(\bar Q q)$ due to the distance separating the $\de$-$\bde$ pair, and thus may remain in this quasi-static state long enough to be detectable as an exotic resonance.  The model then applies the Born-Oppenheimer (BO) approximation to the quasi-static $\de$-$\bde$ pair in order to produce a spectrum of distinct states, as we briefly review here.

Specifically, the Hamiltonian separates into two parts: the kinetic-energy operator $K_{\rm{heavy}}$, referring to the heavy color sources of the system, and the operator $H_{\rm{light}}$, which contains both the interactions between the heavy-source pair and the interaction of the light degrees of freedom (d.o.f.) within the color flux tube.  Explicitly, in the heavy-source center-of-momentum (c.m.) frame,
\begin{equation} 
\label{eq:SepHam}
H=K_{\rm heavy} + H_{\rm light} =
\frac{\mathbf{p}^2}{2 \mu_{\rm heavy}} + H_{\rm light},
\end{equation}
where the full eigenstates may then be written as
\begin{equation} 
\label{eq:AdExp}
|\psi \rangle = \sum_{i} \int d\mathbf{r} \, \tilde \psi_i
(\mathbf{r}) \, |\mathbf{r} \rangle \:
|\xi_i(\mathbf{r}) \rangle .
\end{equation}
Here, $|\mathbf{r} \rangle$ defines the position eigenstate with separation vector $\mathbf r$ between the two heavy color sources.  This separation is in fact general, but exhibits its most transparent interpretation in the context of the so-called \textit{adiabatic expansion}, \textit{i.e.}, in which the state $|\xi (\mathbf{r}) \rangle$ of the light d.o.f.\ depends only upon the instantaneous configuration $|\mathbf{r} \rangle$ of the heavy degrees of freedom.  The set $\left\{ |\xi_i(\mathbf{r}) \rangle \right\}$ forms a basis of eigenstates (labeled by $i$) of $H_{\rm light}$ for any value of $\mathbf{r}$, and $\tilde \psi_i (\mathbf{r})$ are the separation-dependent weights for the individual state components.  Of course, the motivation for performing this expansion lies in the assumed ability to solve the light-field problem outright.  Plugging Eq.~(\ref{eq:AdExp}) back into the \schro equation for the Hamiltonian Eq.~(\ref{eq:SepHam}), one obtains
\begin{equation} \label{eqn:AdiaSchro}
\sum_i \left( - \frac{\hbar^2}{2 \mu_{\de \bde}} [\mathbf{\nabla} + \tau (\mathbf{r})]^2_{ji} + [V_j (\mathbf{r}) - E] \, \delta_{ji}  \right) \! \tilde \psi_i (\mathbf{r}) = 0,
\end{equation}
where we adopt the definition 
\begin{equation}
\mathbf{\tau}_{ji}(\mathbf{r}) \equiv \langle \xi_j (\mathbf{r}) | \nabla \xi_i (\mathbf{r}) \rangle ,
\end{equation}
each of which is called a \textit{Non-Adiabatic Coupling Term\/} (NACT).  One sees that Eq.~(\ref{eqn:AdiaSchro}) diagonalizes if $\mathbf{\tau}_{ji}(\mathbf{r}) \! = \! 0$ for $j \! \neq \! i$, and reduces to a completely conventional \schro equation if the $j \! =\! i$ NACTs vanish as well.

The \textit{adiabatic approximation\/} in this notation implies that if one starts in the $i^{\rm th}$ eigenstate and slightly changes $\mathbf{r}$, then one remains in the original $i^{\rm th}$ eigenstate: $\mathbf{\tau}_{ii}(\mathbf{r}) \! = \! 0$.  The full BO approximation consists of the adiabatic approximation combined with 
\begin{equation} \label{eq:NACT}
\tau_{ji} (\mathbf{r}) = \langle \xi_j (\mathbf{r}) | \nabla \xi_i (\mathbf{r}) \rangle \approx 0,
\end{equation}
for $j \neq i$, which is known as the \textit{single-channel approximation}.  Clearly, this approximation is most accurate for exotic states that lie far in mass from an accessible di-meson threshold to which it may couple, so that the corresponding off-diagonal NACTs are small.

In situations for which this approximation can fail ({\it i.e.}, near thresholds), a rigorous equivalent generalization of the BO approximation is available, the so-called \textit{diabatic formalism}.  First applied to heavy-quark exotics in Ref.~\cite{Bruschini:2020voj} for quarkonium BO states coupling to thresholds, and further developed in Refs.~\cite{Bruschini:2021ckr,Bruschini:2021cty,Lebed:2022vks,Lebed:2023kbm}, this approach allows for the introduction of a coupling between the compact (in our case, $\de$-$\bde$) elementary component and the di-meson thresholds.  This formalism is rigorously equivalent to the NACT approach discussed above~\cite{Baer:2006,Bruschini:2021ckr}, but is more convenient from a computational standpoint.  One begins by replacing Eq.~(\ref{eq:AdExp}) with
\begin{equation} 
\label{eq:DiaExp}
|\psi \rangle = \sum_{i} \int d\mathbf{r}' \tilde \psi_i
(\mathbf{r}' \! , \mathbf{r}_0) \: |\mathbf{r}' \rangle \:
|\xi_i(\mathbf{r}_0) \rangle,
\end{equation}
the \textit{diabatic expansion}.  Crucial to this expansion is the fact that the basis {$|\xi_i(\mathbf{r}) \rangle$} is complete for each value of $\mathbf{r}$, meaning that we are free to choose an arbitrary fiducial $\mathbf{r}_0$ for the basis.  Substituting back into the \schro equation for the Hamiltonian of Eq.~(\ref{eq:SepHam}) yields 
\begin{equation}\label{eq:DiaSchro}
\sum_{i} \left[ - \frac{\hbar^2}{2 \mu_{i}} \de_{ji}  \nabla ^2 +
V_{ji}(\mathbf{r,r_0})-E \de_{ji} \right] \! \tilde \psi_i (\mathbf{r,r_0}) = 0,
\end{equation}
where we have adopted the definition from Ref.~\cite{Bruschini:2020voj} of the \textit{diabatic potential matrix},
\begin{equation}
V_{ji}(\mathbf{r,r_0}) \equiv \langle \xi_j (\mathbf{r}_0)|
H_{\rm light} |\xi_i(\mathbf{r}_0) \rangle,
\end{equation}
the $\mathbf{r}$ dependence appearing through the operator $H_{\rm light}$.  An appropriate choice of separation $\mathbf{r}_0$, \textit{i.e.}, one that is far from any di-meson threshold, directly implies that $|\xi_0(\mathbf{r}_0) \rangle$ can be identified as a pure $\de$-$\bde$ state, and $| \xi_i(\mathbf{r}_0) \rangle$ with $i \! > \! 0$ correspond to pure di-meson states.  It then follows that the diagonal elements of the potential matrix are the static light-field energies of both the pure $\de$-$\bde$ configuration and its associated di-meson configurations, while the off-diagonal elements are mixing terms connecting these configurations.  In the light-field eigenstate basis with $N$ relevant thresholds, $V$ may be expressed in matrix form as 
\begin{equation} \label{eq:FullV}
V=
\begin{pmatrix}
V_{\de \bde}(\mathbf{r}) & V_{\rm mix}^{(1)}(\mathbf{r})  & \cdots &
V_{\rm mix \vphantom{\bbar M_2}}^{(N)}(\mathbf{r}) \\
V_{\rm mix}^{(1)}(\mathbf{r}) & 
V_{M_1 \bbar M_2}^{(1)}(\mathbf r) &
&
\\
\vdots
& & \ddots \\
V_{\rm mix \vphantom{\bbar M_2}}^{(N)}(\mathbf{r}) & & &
V_{M_1 \bbar M_2}^{(N)}(\mathbf r) \\
\end{pmatrix},
\end{equation}
where omitted elements are taken to be zero; {\it i.e.}, we haven taken elements directly connecting distinct di-meson thresholds to be negligible, an approach consistent with analogous calculations in molecular physics (see Ref.~\cite{Baer:2006}).  Following the practice of Refs.~\cite{Bruschini:2020voj,Bruschini:2021ckr,Lebed:2022vks,Lebed:2023kbm}, we set the static light-field energy of each di-meson pair equal to its free energy,
\begin{equation} \label{eq:MMEng}
V_{M_1 \bbar M_2}^{(i)}(\mathbf r) \to T_{M_1 \bbar M_2} = m^{\vphantom\dagger}_{M_1} + m^{\vphantom\dagger}_{M_2} \, .
\end{equation}
While not present in this calculation, a mildly attractive potential may also be added to Eq.~(\ref{eq:MMEng}), roughly modeling rescattering through meson-exchange between the di-meson pair, or through the effects of triangle singularities.

In order to proceed with the calculation, one must now choose a method for handling states whose eigenvalue does not lie near or substantially below the lowest available di-meson threshold.  In Refs.~\cite{Bruschini:2020voj,Lebed:2022vks} a bound-state approximation is applied, in which the authors ignore all thresholds below the eigenstate (to which the state may decay).  In Refs.~\cite{Bruschini:2021ckr,Lebed:2023kbm}, this approximation is lifted in favor of adopting a scattering formalism, where the full suite of available thresholds is included in the potential (and thus the Hamiltonian).  The $S$-matrix may then be calculated by any of the usual means, and a set of normalized cross sections are then extracted, constituting the central results of Refs.~\cite{Bruschini:2021ckr,Lebed:2023kbm}.  However, as indicated above, this approach is not the only available method for handling open thresholds in the diabatic formalism.  As first performed in the diabatic formalism in Ref.~\cite{Bruschini:2021cty}, one may instead use a well-known method~\cite{Fano:1961zz,Eichten:1978tg} to calculate mass corrections and decay widths within the bound-state approximation.  We now review this procedure for the specific application to bound states calculated in the manner presented in Ref.~\cite{Bruschini:2021cty}.

\section{Mass Corrections and Decay Widths From Open Thresholds}\label{sec:MS}

Using much of the notation from the pioneering work of Ref.~\cite{Bruschini:2021cty}, we now present explicit expressions through which the model defined in Sec.~\ref{sec:DDD Model} incorporates the effects of the open thresholds on bound states.  First, one identifies the \textit{interaction Hamiltonian} $H_I$ that connects the elementary bound state $|A\rangle$ to $n$ open thresholds:

\begin{equation} \label{eq:Hint}
H_I(\mathbf{r})=
\begin{pmatrix}
 & V_{\rm mix}^{(1)}(\mathbf{r})  & \cdots &
V_{\rm mix \vphantom{\bbar M_2}}^{(n)}(\mathbf{r}) \\
V_{\rm mix}^{(1)}(\mathbf{r}) & 
 &
&
\\
\vdots
& &  \\
V_{\rm mix \vphantom{\bbar M_2}}^{(n)}(\mathbf{r}) & & &
 \\
\end{pmatrix},
\end{equation}
where again, omitted elements are zero.  For an elementary bound state $|A\rangle$ with quantum numbers $J^{PC}$ interacting with a single open threshold corresponding to a meson pair $B, C$, the shift from its initial mass $M_A$ can be calculated using a standard expression:
\begin{equation} \label{eq:MassShift}
M - M_A = \sum_{m^{\vphantom{2}}_{s_B}, \, m^{\vphantom{2}}_{s_C}} \! \mathcal{P} \! \int \! d \mathbf{p}^{\vphantom\dagger}_{BC} \frac{O_\ell (p^{\vphantom\dagger}_{BC})^2}{M-E_{BC}} \, ,
\end{equation}
where 
\begin{equation} \label{eq:Overlap}
O_\ell (p^{\vphantom\dagger}_{BC}) \equiv {\big|} \langle \mathbf{p}^{\vphantom\dagger}_{BC} ; s^{\vphantom{2}}_B , m_{s_B} , s^{\vphantom{2}}_C , m_{s_C} | H_I | A \rangle {\big|}^2 ,
\end{equation}
and $| \mathbf{p}^{\vphantom\dagger}_{BC} ; s^{\vphantom{2}}_B , m_{s_B} , s^{\vphantom{2}}_C , m_{s_C} \rangle$ is the free di-meson state with c.m.\ momentum $\mathbf{p}^{\vphantom\dagger}_{BC}$.  Additionally, $\mathcal{P}$ indicates the Cauchy principal value operation, which is required in order to handle the singularity present in the denominator.  The free di-meson wave function in position-space representation can be written as
\begin{equation}\label{eq:MM wf}
    \begin{split}
        \langle \mathbf{p}^{\vphantom\dagger}_{BC} ; s_B , m_{s_B} , s_C , m_{s_C} |\mathbf{r} \rangle = \hspace{8em} \\
        \sum_{\ell,m_\ell} \sqrt{\frac{2}{\pi}} \, i^{-\ell} j_\ell (p^{\vphantom\dagger}_{BC} r ) \,
        Y_{\ell}^{m_\ell} (\hat{\mathbf{p}}^{\vphantom\dagger}_{BC}) Y_{\ell}^{m_\ell *} (\hat{\mathbf{r}}) \, \xi^{m_s \dag}_s \, ,
    \end{split}
\end{equation}
where $\mathbf{r}$ indicates the separation vector pointing from $B$ to $C$, $j_\ell$ are spherical Bessel functions, $Y_\ell^{m_\ell}$ are spherical harmonics, and $\xi_s^{m_s}$ are di-meson spin wave functions of total spin $s$.  Using Eq.~(\ref{eq:MM wf}), the overlap in Eqs.~(\ref{eq:MassShift})--(\ref{eq:Overlap}) is
\begin{equation}
\begin{split}
\label{eq:overlap}
\langle \mathbf{p}^{\vphantom\dagger}_{BC} ; s^{\vphantom{2}}_B , m_{s_B} , s^{\vphantom{2}}_C , m_{s_C} | H_I | A \rangle = 
\\
 \sum_{\ell,m_\ell} C^{\, m_{\ell},m_{s},m_J}_{\ell,s,J} \, Y_{\ell}^{m_\ell} (\hat{\mathbf{p}}^{\vphantom\dagger}_{BC}) & \, \mathcal{I}_\ell ( p^{\vphantom\dagger}_{BC} )     ,      
\end{split}
\end{equation}
where $C$ are Clebsch-Gordan coefficients, and
\begin{equation}
\mathcal{I}_\ell ( p_{BC} ) \equiv \sqrt{\frac{2}{\pi}} \, i^{-\ell} \int dr \, r^2 j_\ell (p^{\vphantom\dagger}_{BC} r ) V^{(BC)}_{\rm{mix}} (r) \, u_{\de \bde} (r) ,
\end{equation}
with $V^{(BC)}_{\rm{mix}} (r)$ being the sole off-diagonal element in Eq.~(\ref{eq:Hint}) for the single open-channel case, and $u_{\de \bde} (r)$ is the full radial wave function for the initial elementary $\de$-$\bde$ bound state (which may include contributions from several $\de$-$\bde$ partial waves).  Substituting Eqs.~(\ref{eq:Overlap}) and (\ref{eq:overlap}) back into Eq.~(\ref{eq:MassShift}), and integrating over spherical coordinates, one obtains the mass shift
\begin{equation}
M - M_A = \mathcal{P} \! \int \! dp^{\vphantom\dagger}_{BC} \frac{p_{BC}^2}{M-E_{BC}} \sum_{\ell,s} | \mathcal{I}_\ell ( p^{\vphantom\dagger}_{BC} ) |^2,
\end{equation}
which can be easily rewritten in terms of the $BC$ reduced mass $\mu^{\vphantom{2}}_{BC}$ and a more phenomenologically relevant variable, the free di-meson energy $E^{\vphantom{2}}_{BC}$:
\begin{equation}
M - M_A = \mathcal{P} \! \int \! dE^{\vphantom{2}}_{BC} \, \mu^{\vphantom\dagger}_{BC} \frac{p_{BC}}{M-E_{BC}} \sum_{\ell,s} | \mathcal{I}_\ell ( p^{\vphantom\dagger}_{BC} ) |^2 .
\end{equation}
In this notation, the associated strong decay width $\Gamma$ is
\begin{equation}
    \Gamma_{A \rightarrow BC} =2 \pi \, \mu^{\vphantom\dagger}_{BC} \, p^{\vphantom\dagger}_{BC} \sum_{\ell,s} | \mathcal{I}_\ell ( p^{\vphantom\dagger}_{BC} ) |^2 \Big\rvert _{E^{\vphantom{2}}_{BC \,} = M} \, .
\end{equation}

Generalizing to $n$ open thresholds indexed by $j$, the mass shift is then
\begin{equation}
    M - M_A = \sum^n_{j=1} \mathcal{P} \! \int \! dE_{j} \, \mu_{j} \frac{p_{j}}{M-E_{j}} \sum_{\ell_{j},s_{j}} | \mathcal{I}_{\ell_{j}} ( p_{j} ) |^2 ,
\end{equation}
and the corresponding strong decay width is
\begin{equation} \label{eq:decayformula}
    \Gamma_{A \rightarrow \Sigma_j}=2  \pi \! \sum_{j,\ell_{j},s_{j}} \mu_{j} \, p_{j} \, | \mathcal{I}_{\ell_{j}} ( p_{j} ) |^2 \Big\rvert _{E_{j \,} = M}.
\end{equation}

\section{Results}\label{sec:Results}

Here we present calculated mass corrections and decay widths for the hidden-charm states obtained in Ref.~\cite{Lebed:2023kbm}, which provide the most recent reference points for states first calculated in Ref.~\cite{Lebed:2022vks}.  These results are separated into three flavor sectors ($c\bar c q \bar q^\prime$, $c\bar c q \bar s$, $c\bar c s \bar s$), since the calculations of Refs.~\cite{Lebed:2022vks,Lebed:2023kbm} as well as the ones presented here require the explicit input of diquark (antidiquark) masses.  We take
\begin{eqnarray} \label{eq:demasses}
    m_{\de = (cq)} & = & 1.9271 \ \rm{GeV},
    \nonumber \\
    m_{\de = (cs)} & = & 1.9446 \ \rm{GeV},
\end{eqnarray}
and $m_\de = m_\bde$.  These masses are extracted from two distinct fits: $m_{\de = (cq)}$ is obtained from the most recent fine-structure analysis using the adiabatic dynamical diquark model~\cite{Giron:2021sla}, and $m_{\de = (cs)}$ is obtained from the diabatic-model calculation of Ref.~\cite{Lebed:2023kbm}.  Specifically, $m_{\de = (cq)}$ is obtained by identifying the unique ground-state multiplet isoscalar $J^{PC}=1^{++}$ $c\bar c q\bar q^\prime$ state with $\chi_{c1}{(3872)}$, whose PDG-averaged mass central value is 3871.69~MeV; and $m_{\de = (cs)}$ is obtained by identifying the lowest $\ccss$ state (which has $J^{PC}=0^{++}$) with $X(3915)$, whose PDG-averaged mass is $3921.7$~MeV~\cite{Zyla:2020zbs}.  In addition to these model parameters, we also adopt the same forms for the elementary ($V_{\de \bde}$) and mixing ($V_{\rm mix}^{(i)}$) potentials as in Ref.~\cite{Lebed:2023kbm}, along with their associated parameters:
\begin{equation}
\label{eq:sgmapot}
% Eq. (15) of Lebed:2022vks
V_{\de \bde}(r)=- \frac{\alpha}{r} + \sigma r + V_0 + m_{\de} +
m_{\bde} \, ,
\end{equation}
where $\alpha,\sigma,$ and $V_0$ are given by~\cite{Morningstar:2019}
\begin{eqnarray} \label{eq:sgmapotparams}
\alpha & = & 0.053 \ \rm{GeV} \! \cdot \rm{fm}, \nonumber \\
\sigma & = & 1.097 \ \rm{GeV \! /fm}, \nonumber \\
V_0 & = & -0.180 \ \rm{GeV},
\end{eqnarray}
as well as 
\begin{equation} \label{eq:Mixpot}
% Eq. (16) of Lebed:2022vks
|V_{\rm mix}^{(i)} (r)| = \frac{\Delta}{2}
\exp \! \left\{ -\frac 1 2 \frac{\left[
V^{\vphantom\dagger}_{\de \bde}(r) -
T_{M_1 \bbar M_2 }^{(i)} \right]^2}{(\sigma \rho)^2} \right\} ,
\end{equation}
with $\rho,~\Delta$ being~\cite{Lebed:2023kbm}
\begin{equation} \label{eq:Mxpotparams}
\rho = 0.165 ~\rm{fm}, \ \ 
\Delta = 0.295 ~\rm{GeV}.
\end{equation}
As discussed in Refs.~\cite{Bruschini:2020voj,Bruschini:2021ckr,Bruschini:2021cty,Lebed:2022vks,Lebed:2023kbm}, the particular values of $\rho,\Delta$ adopted for this analysis are not uniquely determined.  Rather, they are one set among many $\rho,\Delta$ combinations that satisfy the conditions of providing both a $J^{PC}=1^{++}$ $\ccqq$ state with mass $3871.69$~MeV and yielding physical behavior for the mixing transition obtained from Eq.~(\ref{eq:Mixpot}).  In addition, while this particular form of mixing potential is phenomenologically motivated by lattice-QCD expectations~\cite{Bulava:2019iut}, it awaits replacement with a fully numerically determined potential to be calculated from detailed lattice-QCD simulations. 

Using the inputs of Eqs.~(\ref{eq:demasses})--(\ref{eq:Mxpotparams}), one may proceed with the methods described in Sec.~\ref{sec:MS} to produce the full set of mass corrections and partial decay widths.  The mass corrections are tabulated in Tables~\ref{tab:ccqq},~\ref{tab:ccss}, and~\ref{tab:ccqs}, while the strong decay widths are collected in Table~\ref{tab:decays}.  In addition to the raw results, we also provide a level diagram in Fig.~\ref{fig:LvlDiagram}, as it is useful to visualize the distribution of available thresholds in each case. 

First, we examine the results for the $\ccqq$ sector, which are presented in Table~\ref{tab:ccqq}.  All states having relevant open thresholds below the previously computed~\cite{Lebed:2022vks} bound-state mass $M_A$ are calculated to have a lower corrected mass $M$, although the shift $M - M_A$ for the $2^{++}$ state is minimal.  When compared to the resonance mass-peak values $M_R$ calculated in Ref.~\cite{Lebed:2023kbm}, we find general agreement for the $0^{++}$ state, $M$ and $M_R$ having a difference of less than 5~MeV, while the $2^{++}$ and $1^{--}$ states have moderate to pronounced ($\sim \!\! 15$--50~MeV) discrepancies, respectively.  Specifically, these calculated masses $M$ lie significantly \textit{above\/} the location of the resonant cross-section peaks $M_R$ obtained in Ref.~\cite{Lebed:2023kbm}. Notably, the $M_R$ value corresponding to the $1^{--}$ state presented here differs from the result one might infer from Ref.~\cite{Lebed:2023kbm}, as here we have also incorporated previously neglected di-meson thresholds (such as $D^* \bar D^*$) that sit quite far below $M_A$.  Thus, we conclude that the methods of Ref.~\cite{Lebed:2023kbm} create a stronger attraction to open thresholds than do the methods of our present work, at least in the $\ccqq$ sector.  As for the disagreement between the locations $M_R$ of the resonant peaks and the calculated shifted masses $M$, one may point to the relatively complex threshold structure of the $2^{++}$ and $1^{--}$ states as compared to the $0^{++}$ state.  That is, since the two methods effectively incorporate each individual threshold differently (in this calculation, through Hamiltonian coupling of the bound state to the free below-threshold di-meson state as in Eq.~(\ref{eq:Overlap}); in Ref.~\cite{Lebed:2023kbm}, through manifestly unitary scattering-theory techniques), then each additional threshold may heighten this effect.  Additionally, we note that while it is useful to compare these results to those for the resonances of Ref.~\cite{Lebed:2023kbm}, which use methods that are the generally more rigorous for handling masses above threshold, the resonant peak values $M_R$ are not precisely the same as the real part of the corresponding pole masses, which arguably serve as a better analogue to the shifted masses $M$ calculated here. 

In the $\ccss$ sector, we also see agreement between the $2^{++}$ states, which have a difference $M - M_R$ of less than 3~MeV\@.  However, the $0^{++}$ and $1^{++}$ mass shifts $M-M_A$ of Table~\ref{tab:ccss} are \textit{positive}, away from the open thresholds.  This result directly contrasts with Ref.~\cite{Lebed:2023kbm}, where the the $0^{++}$ state sees a $\sim \! -6$~MeV shift $M_R - M_A$, and $1^{++}$ a much more dramatic shift of approximately $-45$~MeV\@.  The same substantial discrepancy arises in the $1^{--}$ state as well, even though now $M-M_A$ is small and negative.

Finally, in the $c \bar c q \bar s$ sector, we once again see strong agreement between the $2^{+}$ results, the two having a mere $M - M_R = -2.8$~MeV difference.  The $0^{+}$ state, on the other hand, has an almost exactly opposite shift $M - M_A$, compared to $M_R - M_A = -17.17$~MeV from Ref.~\cite{Lebed:2023kbm}: {\it i.e.}, the $0^+$ mass $M$ of this work is larger than, and the resonant peak mass $M_R$ is smaller than, the elementary bound-state mass $M_A$.  In the $1^{-}$ state first calculated for this work, we find $M-M_R \simeq 20$~MeV\@.  When all three sectors are considered together, a clear trend appears: In each case that $M$ and $M_R$ values show substantial disagreement, $M$ is always greater than $M_R$.  Additionally, the larger discrepancies solely occur within the $J=1$ states.  At bare minimum, these differences suggest that a proper inclusion of spin effects is necessary to obtain precise modeling of the open-threshold effects in a bound-state picture, both in the fine structure of the elementary states, and in the specific di-hadron thresholds~\cite{Bruschini:2023zkb}.  As discussed above, however, one cannot lose sight of the fact that the Hamiltonian and unitary scattering methods incorporate thresholds in two different ways. 

Next, we examine the results of applying Eq.~(\ref{eq:decayformula}) to each state according to valence-quark flavor and $J^{PC}$ content.  One immediately notes the relatively small decay widths throughout Table~\ref{tab:decays}.  For example, one would expect the $0^{++}$ states of both the $\ccqq$ and $\ccss$ sectors to have relatively large widths, especially given their corresponding resonance profiles in the results of Ref.~\cite{Lebed:2023kbm}.  However, we emphasize the decisive effect that our specific choice of mixing potential and parameters [Eqs.~(\ref{eq:Mixpot}) and (\ref{eq:Mxpotparams}), respectively] has on the present calculation.  If a wider mixing potential were applied, for example, one would obtain generally stronger mixing with additional thresholds beyond the closest ones, and consequently larger partial decay widths.  As it stands, many of the states that one might use for comparison of decay rates lack sufficient data for the purpose, \textit{e.g.}, $\psi(4230)$ and its poorly known open-charm decay widths~\cite{ParticleDataGroup:2022pth}.  One could, in fact, work backwards once any such open-flavor partial widths are measured to determine the acceptable ranges for the mixing-potential parameters.

Several other neglected features prevent the current results from being interpreted as the last word in computing the mass shifts and open-flavor decay widths for these states.  First, this particular calculation does not include mixing with conventional quarkonium (although doing so is not difficult~\cite{Lebed:2023kbm}), and such contributions are essential not only for increasing open-flavor decay partial widths, but also provide an avenue for the prominent hidden-flavor decays of these states.    Indeed, our numerical results are not terribly different from those obtained by considering {\em only\/} mixing with quarkonium~\cite{Bruschini:2021cty}.  Second, we have noted (here in Eq.~(\ref{eq:MMEng}) and after, and previously in Refs.~\cite{Lebed:2022vks,Lebed:2023kbm}) that direct interactions between the threshold di-meson pair are neglected but can easily be incorporated into the calculations.  Finally, fine-structure corrections that distinguish states according to isospin and spin, as in the adiabatic calculations of Refs.~\cite{Giron:2019cfc,Giron:2020fvd,Giron:2020qpb,Giron:2020wpx,Giron:2021sla,Giron:2021fnl}, can certainly be included; such effects would have not only direct effects upon the raw spectrum, but would also pull certain states closer to or further away from thresholds, generating pronounced effects upon both the mass spectrum and decay widths.  However, since our purpose in this work has been to perform a preliminary study of the effects on these observables of mixing between $\de$-$\bde$ elementary states and open di-meson thresholds in a landscape of sparse experimental results [{\it e.g.}, in the ground-state multiplet, only $\chi_{c1}(3872)$, $Z_c(3900)$, and $Z_c(4020)$ are clear candidate members], we view the inclusion of such fine-structure corrections here as premature.\footnote{Nevertheless, the fine-structure content does leave an imprint even in the current landscape: As noted in Ref.~\cite{Lebed:2022vks}, the same $D\bar D^*$-$\bar D D^*$ pair in different $C$-eigenstate combinations contributes equally to both $1^{++}$ [$\chi_{c1}(3872)$] and $1^{+-}$ [$Z_c$] channels, and hence the mass difference between these states provides a clear signal of the separate significance of fine structure.}

\begin{table*}[h]
\caption{Calculated mass corrections for the elementary bound states of Ref.~\cite{Lebed:2022vks} with valence quark content $\ccqq$, where $M_A$ ($M$) is the mass eigenvalue calculated excluding (including) mixing with open thresholds.  In addition, we list the peak mass of corresponding scattering resonances in Ref.~\cite{Lebed:2023kbm} as $M_R$.  Note that the $1^{++}$ state has no relevant open thresholds (below $D\bar D^*$, which forms an intrinsic part of the state), and thus we leave its row empty.}
\setlength{\tabcolsep}{9pt}
\renewcommand{\arraystretch}{1.2}
\begin{tabular}{c | c | c | c  } 
 \hline\hline
 $J^{PC}$ & $M-M_A$ (MeV) & $M$ (GeV) & $M_R$ (GeV)
\\
\hline
 $0^{++}$ & $-5.07$ & 3.89876 & 3.89470
 \\
 $1^{++}$ & & 
 \\
 $2^{++}$ & $-0.36$ & 3.91708 & 3.90260
 \\
 $1^{--}$ & $-5.88$ & 4.26370 & 4.21240
 \\
 \hline\hline
\end{tabular}
\label{tab:ccqq}
\end{table*}

\begin{table*}[h]
\caption{The same as Table~\ref{tab:ccqq}, but for valence quark content $\ccss$.}
\setlength{\tabcolsep}{9pt}
\renewcommand{\arraystretch}{1.2}
\begin{tabular}{c | c | c | c  } 
 \hline\hline
 $J^{PC}$ & $M-M_A$ (MeV) & $M$ (GeV) & $M_R$ (GeV)
\\
\hline
$0^{++}$ & \hspace{1em} 2.05 & 3.92374 & 3.91540
 \\
 $1^{++}$ & \hspace{1em} 7.23 & 3.97570 & 3.92230
 \\
 $2^{++}$ & $-19.23$ & 3.93010 & 3.92790
 \\
$1^{--}$ & \hspace{0.2em} $-0.85$ & 4.27787 & 4.22500
 \\
 \hline\hline
\end{tabular}
\label{tab:ccss}
\end{table*}

\begin{table*}[h]
\caption{The same as Table.~\ref{tab:ccqq}, but for valence quark content $c \bar c q \bar s$.  Note that the $1^{+}$ state has no relevant open thresholds (below $D^* \! \bar D_s$-$D \bar D^*_s$, which forms an intrinsic part of the state), and thus we leave its row empty.}
\setlength{\tabcolsep}{9pt}
\renewcommand{\arraystretch}{1.2}
\begin{tabular}{c | c | c | c  } 
 \hline\hline
 $J^{PC}$ & $M-M_A$ (MeV) & $M$ (GeV) & $M_R$ (GeV)
\\
\hline
 $0^{+}$ & \hspace{0.5em} 17.93 & 3.98697 & 3.95190
 \\
 $1^{+}$ & & 
 \\
 $2^{+}$ & $-12.02$ & 3.93940 & 3.94220
 \\
$1^{-}$ & \hspace{0.5em} 23.34 & 4.37086 & 4.35090
 \\
 \hline\hline
\end{tabular}
\label{tab:ccqs}
\end{table*}

\begin{table*}[h]
\caption{Partial and total open-flavor strong decay widths (in MeV) for states computed in this work.} \label{tab:decays}
\renewcommand{\arraystretch}{1.3}
\begin{tabular}{c|c|c|c|c|c|c|c|c|c|c|r}
\hline \hline
    Flavor & $J^{PC}$ & $\Gamma_{D \bar D}$ & $\Gamma_{D \bar D^*}$ & $\Gamma_{D^* \bar D^*}$ & $\Gamma_{D_s \bar D_s}$ &  $\Gamma_{D_{s}^* \bar D_{s}^{*}}$ & $\Gamma_{D \bar D_s}$ & $\Gamma_{D^* \bar D_s}$ & $\Gamma_{D \bar D^*_s}$ & $\Gamma_{D^* \bar D^*_s}$ & $\Gamma_{\rm{total}}$ \\ \hline
     $\ccqq$ & $0^{++}$ & 2.21 & & & & & & & & & 2.21 \\
                  & $2^{++}$ & 0.14 & \ 0.04 & & & & & & & & 0.18 \\
                  & $1^{--}$ & 0.00 & \ 0.04 & 3.80  & & 14.74 & & & & & 18.58 \\
     \hline
     $\ccss$ & $0^{++}$ & 0.90 & & & & & & & & & 0.90 \\
                 & $1^{++}$ & & 24.17 & & & & & & & & 24.17 \\
                 & $2^{++}$ & 4.33 & \ 2.14 & & 0.00 & & & & & & 6.47 \\
                 & $1^{--}$ & 0.00 & \ 0.00 & 1.23 & & 5.57 & & & &  & 6.80 \\
     \hline
     $c \bar c q \bar s$ & $0^{+}$ & & & & & & 6.65 & & & & 6.65 \\
                 & $2^{+}$ & & & & & & 7.23 & & & & 7.23\\
                 & $1^{-}$ & & & & & & 0.67  & 2.13 & 2.14 & 25.56 & 30.50 \\
     \hline \hline
\end{tabular}
\end{table*}

\begin{center}
 
\begin{figure*}

\caption{Shifted mass eigenvalues $M$ of positive-parity states from Tables~\ref{tab:ccqq} and~\ref{tab:ccss}, with relevant di-meson thresholds indicated by dashed lines (where the specific di-meson combination is explicitly labeled) or dotted (for its isospin-partner combination).  The $1^{++}$ $c\bar c q\bar q^\prime$ state, which coincides with $\chi_{c1}(3872)$ and the $D^0 \bar D^{*0}$ threshold, is not shifted by the this calculation.\label{fig:LvlDiagram}}

\textwidth 15.3cm
\textheight 21.9cm
\topmargin 0.0cm
\oddsidemargin 0.1cm
\begin{center}
\begin{tikzpicture}
\def\ncol{3} \def\mmin{3700} \def\mmax{4250} \def\mstep{50}
\def\w{0.33} \def\h{0.5}
\def\mD0{1864.83} \def\mDp{1869.65}
\def\mDstar0{2006.85} \def\mDstarp{2010.26}
\def\mDs{1968.34} \def\mDstars{2112.2} \def\mLambdac{2286.46}
\def\mDOne{2421.1} \def\mDTwo{2461.1}
\datavisualization[scientific axes={width=15.0cm, height=12.0cm},
    x axis={attribute=jpc, label={$J^{PC}$}, min value=\h,
    max value=(\ncol+\h), ticks={step=1,major at={
    1 as $0^{++}$, 2 as $1^{++}$, 3 as $2^{++}$}}},
    y axis={attribute=m, label={Mass (MeV)}, min value=\mmin,
    max value=\mmax, ticks={step=\mstep, minor steps between steps=4,
    style={/pgf/number format/set thousands separator=}}}]
    info {
      \draw [gray, style=dashed]
      (visualization cs: jpc=\h          , m={(\mD0+\mD0)}) --
      (visualization cs: jpc={(\ncol + \h )}, m={(\mD0+\mD0)})
      node [below left,font=\footnotesize] {$D^0 D^0$};
%
  %    \draw [gray, style=dashed]
   %   (visualization cs: jpc={2.5}          , m={(\mD0+\mD0)}) --
   %   (visualization cs: jpc={(\ncol+\h )}, m={(\mD0+\mD0)})
    %  node [below left,font=\footnotesize] {$D^0 D^0$};
%
      \draw [gray, style=dotted]
      (visualization cs: jpc=\h          , m={(\mDp+\mDp)}) --
      (visualization cs: jpc={(\ncol+\h)}, m={(\mDp+\mDp)});
%      node [above left,font=\footnotesize] {$D^+ D^-$};
%
      \draw [gray, style=dashed]
      (visualization cs: jpc=\h          , m={(\mD0+\mDstar0)}) --
      (visualization cs: jpc={(\ncol+\h)}, m={(\mD0+\mDstar0)})
      node [below=1.3ex, left,font=\footnotesize] {$D^0 D^{*0}$};
      \draw [gray, style=dotted]
      (visualization cs: jpc=\h          , m={(\mDp+\mDstarp)}) --
      (visualization cs: jpc={(\ncol+\h)}, m={(\mDp+\mDstarp)});
%      node [above left,font=\footnotesize] {$D^+ D^{*-}$};
%
      \draw [gray, style=dashed]
      (visualization cs: jpc=\h,           m={(\mDs+\mDs)}) --
      (visualization cs: jpc={(\ncol+\h)}, m={(\mDs+\mDs)})
      node [above=1.2ex, left,font=\footnotesize] {$D_s D_s$};
      \draw [gray, style=dashed]
      (visualization cs: jpc=\h,           m={(\mDstar0+\mDstar0)}) --
      (visualization cs: jpc={(\ncol+\h)}, m={(\mDstar0+\mDstar0)})
      node [below=1.3ex, left,font=\footnotesize] {$D^{*0} D^{*0}$};
      \draw [gray, style=dotted]
      (visualization cs: jpc=\h,           m={(\mDstarp+\mDstarp)}) --
      (visualization cs: jpc={(\ncol+\h)}, m={(\mDstarp+\mDstarp)});
%      node [above left,font=\footnotesize] {$D^{*+} D^{*-}$};
%
      \draw [gray, style=dashed]
      (visualization cs: jpc=\h, m={(\mDs+\mDstars)}) --
      (visualization cs: jpc={(\ncol+\h)}, m={(\mDs+\mDstars)})
      node [below=1.5ex,left,font=\footnotesize] {$D_s D_s^*$};
      \draw [gray, style=dashed]
      (visualization cs: jpc=\h, m={(\mDstars+\mDstars)}) --
      (visualization cs: jpc={(\ncol+\h)}, m={(\mDstars+\mDstars)})
      node [below=1.5ex, left,font=\footnotesize] {$D_s^* D_s^*$};
	 \draw [black]
	 (visualization cs: jpc={(1-\w)}, m={(3899)})
	 rectangle 
	 (visualization cs: jpc={(1+\w)}, m={(3899)})
	 node [below=0.0ex, left=11ex, mark size=3pt] {\pgfuseplotmark{triangle}};
	 \draw [black]
	 (visualization cs: jpc={(1-\w)}, m={(3924)})
	 rectangle
	 (visualization cs: jpc={(1+\w)}, m={(3924)})
	 node [below=0.0ex, left=11ex] {\pgfuseplotmark{square}};
	 \draw [black]
	 (visualization cs: jpc={(2-\w)}, m={(3976)})
	 rectangle
	 (visualization cs: jpc={(2+\w)}, m={(3976)})
	 node [below=0.0ex, left=11ex] {\pgfuseplotmark{square}};
	 \draw [black]
	 (visualization cs: jpc={(3-\w)}, m={(3917)})
	 rectangle
	 (visualization cs: jpc={(3+\w)}, m={(3917)})
	 node [below=0.0ex, left=11ex, mark size=3pt] {\pgfuseplotmark{triangle}};
	 \draw [black]
	 (visualization cs: jpc={(3-\w)}, m={(3930)})
	 rectangle
	 (visualization cs: jpc={(3+\w)}, m={(3930)})
	 node [below=0.0ex, left=11ex] {\pgfuseplotmark{square}};
	\matrix [draw, above left] at (14,9.5) {
  	\node[mark size=3pt] {\raisebox{0.5ex}{\pgfuseplotmark{triangle}} $\, = c\bar c q \bar  q^\prime$}; \\
  	\node[left=0.70ex] {\raisebox{0.5ex}{\pgfuseplotmark{square}} $\, = c\bar c s \bar s$}; \\
	};  
    };
\end{tikzpicture}
\end{center}
\end{figure*}
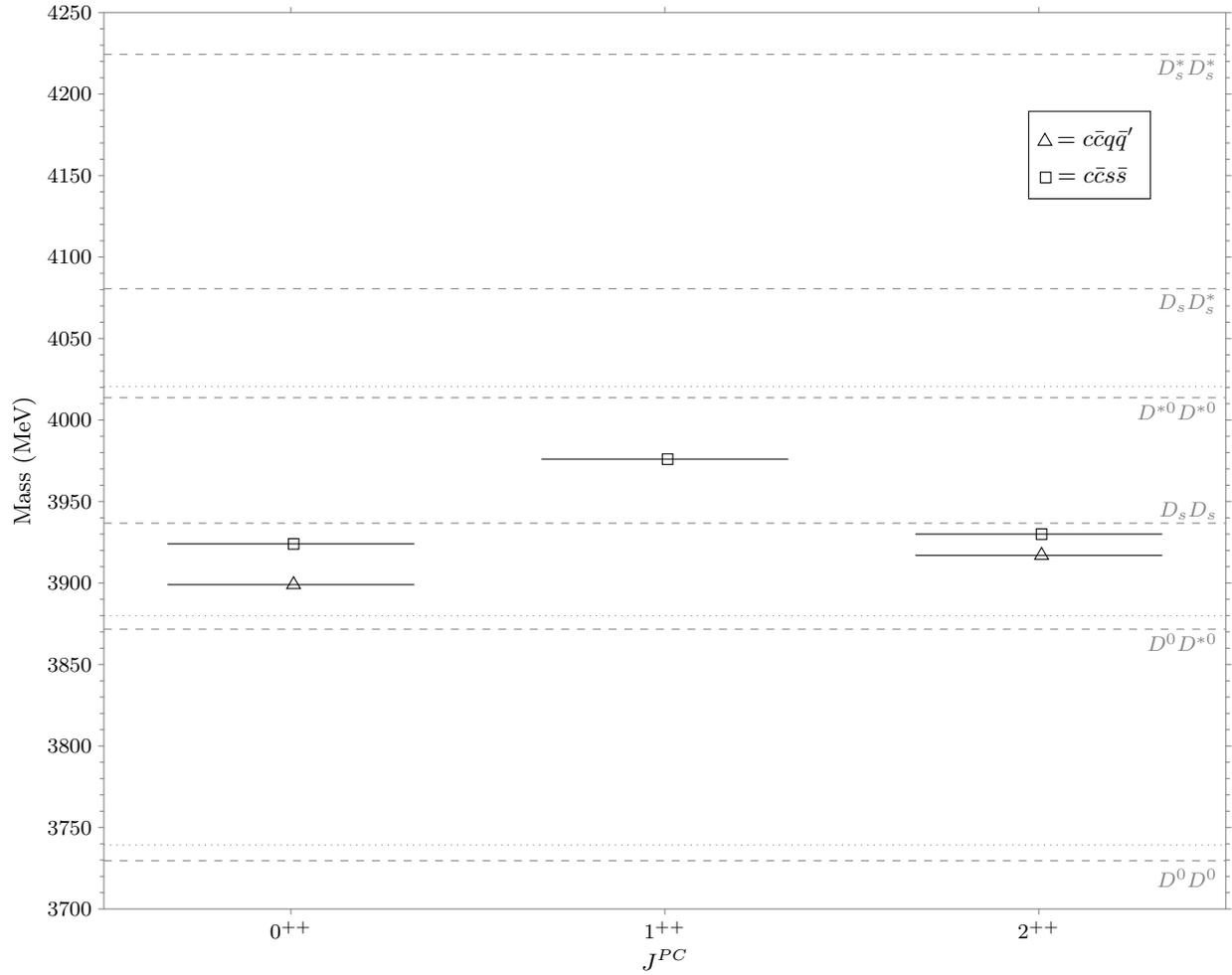
   
\end{center}

\section{Conclusions}\label{sec:Conclusions}

The diabatic extension~\cite{Lebed:2022vks,Lebed:2023kbm} to the adiabatic dynamical diquark model~\cite{Brodsky:2014xia,Lebed:2017min} rigorously extends the Born-Oppenheimer approximation inherent in the original model to allow for the inclusion of effects of mixing with nearby di-hadron thresholds.  In this work, we use well-known Hamiltonian techniques~\cite{Eichten:1978tg,Fano:1961zz} for calculating the effect of open di-hadron thresholds that lie significantly below the original bound state.  We compute mass shifts and open-flavor partial decay widths due to the presence of these thresholds in multiple ($c\bar c q\bar q^\prime$, $c\bar c s\bar s$, $c\bar c q \bar s$) flavor sectors, and find that in most cases the mass eigenvalues lie quite close to (within a few MeV) those obtained from a fully unitarized scattering calculation~\cite{Lebed:2023kbm}.  In the cases where a substantial discrepancy arises, we attribute the difference to the requirement of unitarity imposing constraints between the elementary $\de$-$\bde$ state and the allowed thresholds.

We also find that the open-flavor partial decay widths, while broadly occurring in physically acceptable ranges, might be viewed as surprisingly small in many cases [{\it e.g.}, $O(100~{\rm keV})$].  Absent direct input from experimental measurements, such results may in fact turn out to be entirely physical; or at minimum, the discrepancies will point the way to understanding the precise nature of the threshold mixing potential.

This work serves as another milepost for developing the diabatic dynamical diquark model into a complete formalism capable of addressing the entire sector of heavy-quark exotic hadrons.  However, it should also be noted that the same techniques apply to any system for which mixing with elementary states thresholds is an essential feature, such as hybrids and glueballs.  In the diquark model, the next major step for future work is a study of the effect of including direct rescattering interactions between the di-meson pair.  Other planned improvements include the incorporation of fine-structure (spin and isospin) corrections into the elementary states, and spin corrections to the di-meson threshold states.

%\vspace{-2ex}
\begin{acknowledgments}
This work was supported by the National Science Foundation (NSF) under 
Grants No.\ PHY-1803912 and PHY-2110278.
\end{acknowledgments}

\bibliographystyle{apsrev4-2}
\bibliography{diquark}
\end{document}